\documentclass[conference]{IEEEtran}

\IEEEoverridecommandlockouts

\usepackage{amsthm, amssymb}

\def\b{\mathbb}
\def\bb{\boldsymbol}

\newtheoremstyle{slplain}
  {3pt}
  {3pt}
  {\slshape}
  {}
  {\bfseries}
  {.}%
  { }
  {}

\theoremstyle{slplain}

\def\bb{\boldsymbol}
\def\b{\mathbb}

\def\bb{{\mathbf{b}}}

\def\bee{{\mathbf{e}}}

\def\b0{{\mathbf{0}}}

\def\bA{{\mathbf{A}}}

\usepackage{cite}

\usepackage{amsfonts}
\usepackage{multicol,multienum}
\usepackage{subfigure}

\ifCLASSINFOpdf
  \usepackage[pdftex]{graphicx}
  \graphicspath{{../eps/}{../jpeg/}}
  \DeclareGraphicsExtensions{.eps}
\else
  \usepackage{float}
  \usepackage[dvips]{graphicx}
  \graphicspath{{../eps/}}
  \DeclareGraphicsExtensions{.eps}
\fi

\usepackage[cmex10]{amsmath}
\usepackage{algorithm}
\usepackage{algorithmic}
\usepackage{array}
\usepackage{setspace}
\usepackage{url}

\begin{document}

\title{Doppler Shift Estimation in 5G New Radio Non-Terrestrial Networks}

\author{
\IEEEauthorblockA{Xingqin Lin, Zhipeng Lin, Stefan Eriksson Löwenmark, Johan Rune, and Robert Karlsson}
\IEEEauthorblockA{Ericsson}
\IEEEauthorblockA{Contact: xingqin.lin@ericsson.com}
\thanks{\copyright 2021 IEEE.  Personal use of this material is permitted.  Permission from IEEE must be obtained for all other uses, in any current or future media, including reprinting/republishing this material for advertising or promotional purposes, creating new collective works, for resale or redistribution to servers or lists, or reuse of any copyrighted component of this work in other works.
}
}

\maketitle
%\doublespacing
\thispagestyle{empty}

\begin{abstract}
Evolving 5G New Radio (NR) to support non-terrestrial networks (NTNs), particularly satellite communication networks, is under exploration in 3GPP. The movement of the spaceborne platforms in NTNs may result in large timing varying Doppler shift that differs for devices in different locations. Using orthogonal frequency-division multiple access (OFDMA) in the uplink, each device will need to apply a different frequency adjustment value to compensate for the Doppler shift. To this end, the 3GPP Release-17 work on NTNs assumes that an NTN device is equipped with a global navigation satellite system (GNSS) chipset and thereby can determine its position and calculate the needed frequency adjustment value using its position information and satellite ephemeris data. This makes GNSS support essential for the NTN operation. However, GNSS signals are weak, not ubiquitous, and susceptible to interference and spoofing. We show that devices without access to  GNSS signals can utilize reference signals in more than one frequency position in an OFDM carrier to estimate the Doppler shift and thereby determine the needed frequency adjustment value for pre-compensating the Doppler shift in the uplink. We analyze the performance, elaborate how to utilize the NR reference signals, and present simulation results. The solution can reduce the dependency of NTN operation on GNSS with reasonable complexity and performance trade-off.
\end{abstract}

\IEEEpeerreviewmaketitle

\section{Introduction}
\label{sec:intro}

New Radio (NR) is the fifth-generation (5G) wireless access technology that addresses a wide range of use cases from enhanced mobile broadband (eMBB) to ultra-reliable low-latency communications (URLLC) to massive machine type communications (mMTC) [1]. After finalizing the first 5G NR specifications in the 3rd generation partnership project (3GPP) Release 15, 3GPP has been working on the evolution of 5G NR technology to improve performance and address new use cases. Evolving 5G NR to support non-terrestrial networks (NTNs) is under exploration in 3GPP [2]. NTN can be used as an umbrella term for any network that involves non-terrestrial flying objects. The focus of the 3GPP NTN work has been satellite communication networks utilizing spaceborne platforms which include low Earth orbiting (LEO) satellites, medium Earth orbiting (MEO) satellites, and geosynchronous Earth orbiting (GEO) satellites.

The movement of the spaceborne platforms in NTN may result in large Doppler shifts. For example, the high speed of a LEO satellite at the height of 600 km can lead to a location-dependent and time-varying Doppler shift as large as 24 ppm (amounting to 48 kHz at 2 GHz carrier frequency) [3]. To access an NR network, a user equipment (UE) must acquire time and frequency synchronization in the downlink (DL) using synchronization signals (SS). UEs in the same cell may tune to significantly different frequencies due to different Doppler shifts. If the frequency of the DL signal is used as frequency reference for the UL, the DL Doppler shift will translate to a corresponding frequency shift in the UL at the UE. In addition, the UL signals will also be subjected to Doppler shifts. This is illustrated in Figure 1, where one UE experiences a positive Doppler shift while the other a negative Doppler shift. The uplink signals are transmitted at frequencies misaligned by the DL Doppler shift difference and received by base station (BS) at frequencies further misaligned by the UL Doppler shift difference. Note that NR adopts orthogonal frequency division multiplexing (OFDM) waveform and its UL multi-access scheme is based on orthogonal frequency division multiple access (OFDMA). Due to the frequency misaligned uplink transmissions as illustrated in Figure 1, the orthogonality of OFDMA would be significantly impacted.

\begin{figure}
\centering
\includegraphics[width=7cm]{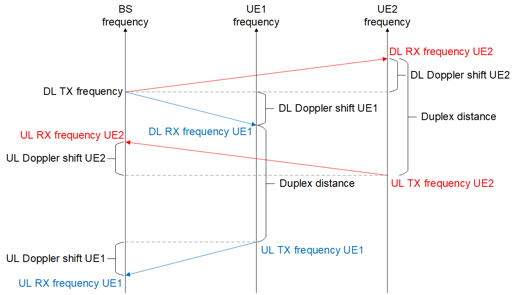}
\caption{An illustration of frequency misaligned uplink transmissions from two UEs in the presence of Doppler shifts, assuming BS is at the satellite.}
\label{fig:1}
\end{figure}

To align the received frequency of the uplink transmissions, each UE needs to apply a different frequency adjustment in the uplink to compensate for its specific Doppler shift. Then, the central question is how to determine the needed frequency adjustment at each UE. In the 3GPP Release 17 work on NTNs, the assumption is that each NTN UE is equipped with a global navigation satellite system (GNSS) chipset and can determine its position and velocity [4]. The NTN can periodically broadcast satellite ephemeris data which conveys information about the serving satellite’s position and velocity. Therefore, with the known positions and velocities of the satellite and UE, the UE can estimate the DL and UL Doppler shifts. Prior to initial access, the UE can adjust its UL transmit frequency by the sum of the estimated UL and DL Doppler shifts (in opposite direction of the Doppler shift). This helps achieve multi-access orthogonality at the BS receiver.

The GNSS signals are weak, not ubiquitous, and susceptible to interference and spoofing [5]. As a result, there are situations where an NTN UE temporarily lacks proper GNSS coverage. A typical situation may be when the UE is inside a building, but this may occur in other situations too. During periods of temporary loss of GNSS coverage, a UE may not be able to accurately pre-compensate the Doppler shift and the UE may not receive the services normally provided by the NTN to the UE. To mitigate the consequence of lack of GNSS coverage or to support NTN UEs without GNSS support, it is desirable to enable the UE to estimate the Doppler shift without GNSS support [6]. The frequency error the UE observes in the DL is a combination of Doppler shift (mainly due to satellite movement) and frequency offset (due to crystal oscillator mismatch between the transmitting satellite and the receiving UE), making it difficult to separate the two effects. Despite there exists rich literature on frequency offset estimation for OFDM transmission (see, e.g. [7][8][9]) as well as Doppler shift estimation (see, e.g. [10][11]), these methods cannot distinguish between the large Doppler shift in NTN and the frequency offset caused by a crystal oscillator. 

As there are two unknowns (Doppler shift and frequency offset), we need to measure at least two DL signals on different frequencies to estimate both. To this end, we utilize reference signals in more than one frequency position in an OFDM carrier to jointly estimate Doppler shift and frequency offset and then separate the two effects. With the estimated Doppler shift, a UE can determine the needed frequency adjustment value and apply it to frequency pre-compensation of UL transmissions so that multi-access orthogonality at the receiver is maintained. The advantages of the proposed solution include mitigating the consequences of loss of GNSS coverage for a UE operating in an NTN, enabling a UE without GNSS capability to access an NTN, reducing the dependency of NTN operation on GNSS, among others.

The remainder of this article is organized as follows. In Section II, we explain the design rationale by describing the signal processing model and analysis. In Section III, we elaborate how the NR reference signals can be exploited to facilitate the joint estimation of Doppler shift and frequency offset. Simulation results are presented in Section IV, followed by our concluding remarks in Section V.

\section{Doppler Shift Estimation}
\label{sec:doppler}

\subsection{Baseband model}

In this section, we describe the baseband representation of OFDM signal transmission and reception to elaborate the basic idea.

Consider an OFDM system with $W$ Hz bandwidth and $B$ Hz subcarrier spacing. The discrete Fourier transform (DFT) size, denoted as $N$, is usually chosen as a power of 2 larger than the number $W/B$ of subcarriers. Denote by $X[k$] the symbol on subcarrier $k$ and $N_{cp}$ the length of cyclic prefix. The baseband equivalent transmit signal can then be written as
\begin{align}
x[n] = \frac{1}{N} \sum_{k=0}^{N-1} X[k] e^{j2\pi\frac{n}{N}k}, n = - N_{cp}, ..., N-1.
\end{align}

For the line-of-sight dominated NTN wireless channel, the channel can be modeled as a one-tap channel, while in the simulation results presented in Section IV, we use the NTN tapped delay line (TDL) models with multiple taps [13]. Denoting by $h[n]$ the channel coefficient at the $n$-th time-domain sample:
\begin{align}
h[n] = \bar{h} e^{j2\pi\frac{v}{\lambda}nT_s}
\end{align}
where $\bar{h}$ is the complex channel coefficient without Doppler shift, $\lambda$ is the carrier wavelength, $v$ is the relative speed between the transmitter and the receiver, and $T_s$ is the sampling interval equal to $1/NB$. The baseband equivalent received signal is then given by
\begin{align}
y[n] &= e^{j2\pi \Delta f nT_s} e^{j2\pi\frac{v}{\lambda}nT_s}   \bar{h} x[n] + w[n] \notag \\
&=  e^{j2\pi  \left( \Delta f + \frac{v}{\lambda} \right ) nT_s}   \bar{h} x[n] + w[n] 
\end{align}
where $\Delta f $ is the frequency offset due to the oscillator frequency mismatch between the transmitter and the receiver, and $w[n]$ denotes noise samples. 

The receiver discards the first $N_{cp}$  samples and performs $N$-point DFT operation of the samples $n=0,…,N-1$, which yields the received symbols in frequency domain as
\begin{align}
&Y[k] =  \sum_{n=0}^{N-1} y[n] e^{-j2\pi\frac{k}{N}n}  \notag \\
%&=  \sum_{n=0}^{N-1} e^{j2\pi  \left( \Delta f + \frac{v}{\lambda} \right ) nT_s}   \bar{h} x[n] e^{-j2\pi\frac{k}{N}n} +  \sum_{n=0}^{N-1} w[n] e^{-j2\pi\frac{k}{N}n}  \notag \\
&= \bar{h}  \sum_{n=0}^{N-1}  x[n] e^{-j2\pi \frac{k- \frac{1}{B}\left( \Delta f + \frac{v}{\lambda} \right ) }{N} n}       + W[k] \notag \\
&= \bar{h}  \sum_{n=0}^{N-1}  \frac{1}{N} \sum_{\ell=0}^{N-1} X[\ell] e^{j2\pi\frac{n}{N}\ell} e^{-j2\pi \frac{k- \frac{1}{B}\left( \Delta f + \frac{v}{\lambda} \right ) }{N} n}       + W[k] \notag \\
&= \bar{h} \sum_{\ell=0}^{N-1}   \frac{1}{N} X[\ell]   \sum_{n=0}^{N-1}    e^{j2\pi \frac{\ell - k + \frac{1}{B}\left( \Delta f + \frac{v}{\lambda} \right ) }{N} n}       + W[k]
\end{align}
where $W[k]$ denotes the frequency-domain noise. Note that
\begin{align}
  &\sum_{n=0}^{N-1}    e^{j2\pi \frac{\ell - k + \frac{1}{B}\left( \Delta f + \frac{v}{\lambda} \right ) }{N} n}   = \frac{1- e^{j2\pi \left( \ell - k + \frac{1}{B}\left( \Delta f + \frac{v}{\lambda} \right ) \right) } }{1 - e^{j2\pi \frac{\ell - k + \frac{1}{B}\left( \Delta f + \frac{v}{\lambda} \right ) }{N}}} = \notag \\
 &  e^{j\pi \left( \ell - k + \frac{1}{B}\left( \Delta f + \frac{v}{\lambda} \right ) \right) \left( 1 - \frac{1}{N} \right) } \frac{\sin \left( \pi \left( \ell - k + \frac{1}{B}\left( \Delta f + \frac{v}{\lambda} \right ) \right)  \right)}{\sin \left( \pi \left( \ell - k + \frac{1}{B}\left( \Delta f + \frac{v}{\lambda} \right ) \right)/N  \right)}.
\end{align}
Thus, plugging (5) into (4) yields that
\begin{align}
Y[k] =&  \frac{\bar{h}}{N} \sum_{\ell=0}^{N-1}    \frac{\sin \left( \pi \left( \ell - k + \frac{1}{B}\left( \Delta f + \frac{v}{\lambda} \right ) \right)  \right)}{\sin \left( \pi \left( \ell - k + \frac{1}{B}\left( \Delta f + \frac{v}{\lambda} \right ) \right)/N  \right)} \cdot \notag \\
&e^{j\pi \left( \ell - k + \frac{1}{B}\left( \Delta f + \frac{v}{\lambda} \right ) \right) \left( 1 - \frac{1}{N} \right) } X [\ell] + W[k] .
\end{align}

If there is no Doppler shift $\frac{v}{\lambda}$ and the frequency offset $\Delta f$ is zero, (6) reduces to the form:
\begin{align}
Y[k] =&  \frac{\bar{h}}{N} \sum_{\ell=0}^{N-1}    \frac{\sin \left( \pi \left( \ell - k   \right)  \right)}{\sin \left( \pi \left( \ell - k  \right)/N  \right)} e^{j\pi \left( \ell - k \right) \left( 1 - \frac{1}{N} \right) } X [\ell] + W[k] \notag \\
=& \bar{h} X[k] + W[k].
\end{align}
This is the canonical form of OFDM transmission: The data symbols modulate parallel subcarriers and they are maintained to be orthogonal as they propagate through the channel. With the presence of the Doppler shift $\frac{v}{\lambda}$ and the frequency offset $\Delta f$, as shown in (6), there is inter-subcarrier interference between the symbols transmitted on different subcarriers. Thus, accurate carrier synchronization is essential for OFDM systems.

\subsection{Receiver algorithm}

In this section, we describe the receiver algorithm for estimating the Doppler shift $\frac{v}{\lambda}$ and the frequency offset $\Delta f$. To this end, assume that $X[k]$ are a known reference signal, and accordingly $x[n]$ is also known. The receiver can then perform element-wise multiplication of the received signal $y[n]$ with the transmitted signal $x[n]$, which yields the sequence $z[n]$ given by 
\begin{align}
z[n] &= x^*[n] y [n] \notag \\
&=  e^{j2\pi  \left( \Delta f + \frac{v}{\lambda} \right ) nT_s}    \bar{h} |x[n]|^2 + x^*[n] w[n]
\end{align}
where $x^*[n]$ is the conjugate of $x[n]$.

To remove the unknown phase rotation introduced by the complex channel $\bar{h}$, the receiver can apply a differential operation to the sequence $z[n]$, which yields that
\begin{align}
&z^*[n-D] z[n] =  e^{j2\pi  \left( \Delta f + \frac{v}{\lambda} \right ) D T_s}   |\bar{h}|^2 |x[n]|^2 |x[n-D]|^2 \notag \\
&+ e^{-j2\pi  \left( \Delta f + \frac{v}{\lambda} \right ) (n-D)T_s}    \bar{h}^* |x[n-D]|^2 x^*[n] w[n] \notag \\
&+e^{j2\pi  \left( \Delta f + \frac{v}{\lambda} \right ) nT_s}    \bar{h} |x[n]|^2  x[n-D] w^*[n-D] \notag \\
&+  x[n-D] x^*[n] w^*[n-D] w[n]
\end{align}
where $D$ is a design parameter that determines the step size of the differential operation. Discarding the noise terms and summing up $z^*[n-D] z[n]$ over $n$, the receiver can form the estimation equation:
\begin{align}
 \Delta f + \frac{v}{\lambda} = \frac{1}{2\pi D T_s} \cdot \textrm{Phase} \left( \sum_n z^*[n-D] z[n]   \right).
\end{align}

Due to the periodicity of the discrete-time exponential, the following condition needs to be satisfied to make the estimate accurate:
\begin{align}
 \Delta f + \frac{v}{\lambda} \leq \frac{1}{2D T_s} .
\end{align}

There is a tradeoff for the estimation. Choosing a larger $D$ improves the estimate since it results in more noise averaging. However, choosing a larger $D$ reduces the range of offsets that can be estimated.

This method can estimate the composite effect of the Doppler shift $\frac{v}{\lambda}$ and the frequency offset $\Delta f$, but it cannot separate the two effects. To solve for the two unknowns, we need more than one estimation equation. The basic idea is to utilize known reference signals at more than one frequency position in the OFDM carrier. To this end, denote by $\{ X[k] \}_{k \in S_p}$  the $p$-th reference signal in the bandwidth of the OFDM carrier, where $S_p$ denotes the set of subcarriers used to transmit the $p$-th reference signal. The center frequency of the reference signal is denoted as $f_c+f_p$, where $f_c$ is the carrier frequency and $f_p$ denotes the offset of the center frequency of the reference signal relative to the carrier frequency $f_c$. The corresponding time domain reference signal is denoted as $x_p [n]$.

The receiver filters the received signal to extract out the corresponding signal component for the $p$-th reference signal, which is denoted as $y_p [n]$. Then the receiver can perform element-wise multiplication of the received signal $y_p [n]$ with the transmitted reference signal $x_p [n]$ to produce $z_p [n] = x_p^* [n] y_p [n]$. Accordingly, the receiver can form the estimation equation:
\begin{align}
\Delta f + \frac{v}{c} (f_c + f_p) =  \frac{1}{2\pi D T_s} \cdot \textrm{Phase} \left( \sum_n z^*_p[n-D] z_p[n]   \right)
\end{align}
where $c$ denotes the speed of light. Denote by $P$ the number of frequency positions used for transmitting the known reference signals. Accordingly, we have $P$ estimation equations. The system of linear equations can be expressed in matrix form as
\begin{align}
\bA \bee = \bb
\end{align}
where
\begin{align}
\bee &= 
\begin{bmatrix}
\Delta f \\
v 
\end{bmatrix} \\
\bb &= \begin{bmatrix}
 \frac{1}{2\pi D T_s} \cdot \textrm{Phase} \left( \sum_n z^*_1[n-D] z_1[n]   \right) \\
 \frac{1}{2\pi D T_s} \cdot \textrm{Phase} \left( \sum_n z^*_2[n-D] z_2[n]   \right) \\
\vdots   \\
 \frac{1}{2\pi D T_s} \cdot \textrm{Phase} \left( \sum_n z^*_P[n-D] z_P[n]   \right) \\
\end{bmatrix} \\
\bA &= \begin{bmatrix}
1 & \frac{f_c + f_1}{c} \\
1 & \frac{f_c + f_2}{c} \\
\vdots & \vdots \\
1 & \frac{f_c + f_P}{c} 
\end{bmatrix}
\end{align}
Since we assume that $P \geq 2$, $\bA$ is a tall matrix and the system is overdetermined in general. In this case, we can pursue an approximate solution known as least squares (LS) given by
\begin{align}
\bee_{LS} = ( \bA^\dagger \bA )^{-1} \bA^\dagger \bb.
\end{align}
where $\bA^\dagger$ denotes the Hermitian transpose of $\bA$. Furthermore, the estimate can be filtered over time, for example, by a first order infinite impulse response (IIR) filter:
\begin{align}
\begin{bmatrix}
\tilde{\Delta f} [m] \\
\tilde{v} [m] 
\end{bmatrix} = (1 - \gamma) \begin{bmatrix}
\tilde{\Delta f} [m-1] \\
\tilde{v} [m-1] 
\end{bmatrix}  + \gamma \bee_{LS} [m]
\end{align}
where $[\tilde{\Delta f} [m] \ \tilde{v} [m] ]^T$ is the filtered estimation at the $m$-th instant, $\gamma \in [0, 1]$ is a filter parameter, and $\bee_{LS} [m]$ is the unfiltered estimation at the $m$-th instant.

\subsection{Impact of timing drift}

In the prevision section, we use the nominal sample rate $f_s=1/T_s$ in the analysis. Denote by $\alpha = f_s/f_c$ the ratio of the nominal sampling rate and the carrier frequency. The actual sampling rate is then given by $\alpha(f_c-\Delta f)$. Taking this effect into account, the estimation equation would become
\begin{align}
\Delta f + \frac{v}{c} (f_c + f_p) =  \frac{\textrm{Phase} \left( \sum_n z^*_p[n-D] z_p[n]   \right)}{2\pi D T_s^{'}}
\end{align}
where $T_s^{'} = \frac{1}{\alpha(f_c-\Delta f)}$ denotes the actual sampling interval at the receiver. Plugging the actual sampling interval in the equation yields
\begin{align}
\Delta f + \frac{v}{c} (f_c + f_p) =  \frac{\alpha(f_c-\Delta f) \cdot \textrm{Phase} \left( \sum_n z^*_p[n-D] z_p[n]   \right) }{2\pi D}
\end{align}
Rearranging the equation yields
\begin{align}
&\left( 1 +   \frac{\alpha \cdot \textrm{Phase} \left( \sum_n z^*_p[n-D] z_p[n]   \right) }{2\pi D}  \right) \Delta f + \frac{v}{c} (f_c + f_p)  \notag \\
&=  \frac{\alpha f_c }{2\pi D}   \cdot \textrm{Phase} \left( \sum_n z^*_p[n-D] z_p[n]   \right) \notag \\
&= \frac{1 }{2\pi D T_s} \cdot \textrm{Phase} \left( \sum_n z^*_p[n-D] z_p[n]   \right) .
\end{align}

The system of linear equations can be expressed in matrix form as
\begin{align}
\bA{'} \bee = \bb
\end{align}
where
\begin{align}
\bA{'} &= \begin{bmatrix}
1 + \frac{\alpha \cdot \textrm{Phase} \left( \sum_n z^*_1[n-D] z_1[n]   \right) }{2\pi D} & \frac{f_c + f_1}{c} \\
1 + \frac{\alpha \cdot \textrm{Phase} \left( \sum_n z^*_2[n-D] z_2[n]   \right) }{2\pi D} & \frac{f_c + f_2}{c} \\
\vdots & \vdots \\
1 + \frac{\alpha \cdot \textrm{Phase} \left( \sum_n z^*_P[n-D] z_P[n]   \right) }{2\pi D} & \frac{f_c + f_P}{c} 
\end{bmatrix} .
\end{align}
This system of linear equations can be similarly solved using the LS algorithm.

Note that the ratio of the nominal sampling rate and the carrier frequency is typically much smaller than 1, i.e., $\alpha = f_s/f_c \ll 1$. Therefore,  
\begin{align}
\frac{\alpha \cdot \textrm{Phase} \left( \sum_n z^*_p[n-D] z_p[n]   \right) }{2\pi D} \leq \frac{\alpha}{D} \leq \alpha \ll 1.
\end{align}
In other words, 
\begin{align}
1 + \frac{\alpha \cdot \textrm{Phase} \left( \sum_n z^*_p[n-D] z_p[n]   \right) }{2\pi D} \approx 1.
\end{align}
Accordingly, $\bA \approx \bA{'}$. Thus, the impact of timing drift is negligible in the estimate of Doppler shift and frequency offset.

\section{Reference Signals}
\label{sec:signal}

As described in Section II, a network node transmits reference signals in more than one frequency position in the OFDM carrier to enable a UE to jointly estimate Doppler shift and frequency offset and then separate the two effects. There are many different types of reference signals in NR [14]. In the sequel, we describe one example design based on downlink synchronization signals. 

To access an NR network, a UE needs to carry out initial-access functionality which includes cell search and random access. To enable the UE to acquire DL time and frequency synchronization, an SS consisting of the primary SS (PSS) and the secondary SS (SSS) is periodically transmitted in the DL of each cell. After synchronization, the UE can decode the physical broadcast channel (PBCH) which carries the master information block (MIB) that the UE needs to decode in order to receive the remaining system information broadcast by the network. In NR, the PSS, SSS, and PBCH are jointly referred to as SS block (SSB) which occupies 20 resource blocks (RBs). After decoding the PBCH, the UE can move forward to decode the system information block type 1 (SIB1) which contains the system information that the UE needs to know before accessing the network. For example, SIB1 contains information about random access configuration that the UE needs in order to carry out random access procedure. Since the SSB has an associated SIB1 transmission, it is referred to as cell-defining SSB (CD-SSB).

The first step of the random access procedure is the UE's transmission of a random access preamble in the uplink. The UE needs to determine the uplink frequency adjustment value and apply frequency pre-compensation (and timing advance) before transmitting the random access preamble. We propose that the network broadcasts the time and frequency locations of at least one additional SSB in SIB1. The additional SSB does not have an associated SIB1 transmission and thus is referred to as a non-cell-defining SSB (non-CD-SSB). 

With the CD-SSB and at least one non-CD-SSB, the UE can use the method described in Section II to estimate the Doppler shift and apply the corresponding frequency pre-compensation in its UL transmissions including the random access preamble transmission. Figure 2 provides an illustration of the UE procedure in utilizing two SSBs located in different frequency positions for the joint estimation of Doppler shift and frequency offset. To facilitate accurate estimation, it is preferable that the two SSBs are located far apart in the frequency domain, e.g., at opposite ends of frequency band or with the CD-SSB at the center of the band and the non-CD-SSB at one of the edges of the frequency band. The network could also broadcast one non-CD-SSB at each end of the frequency band, in order to have maximum frequency difference between the SSBs that the UE uses for the joint estimation of Doppler shift and frequency offset, while still allowing the CD-SSB to be located at the center of the frequency band.

\begin{figure}
\centering
\includegraphics[width=6cm]{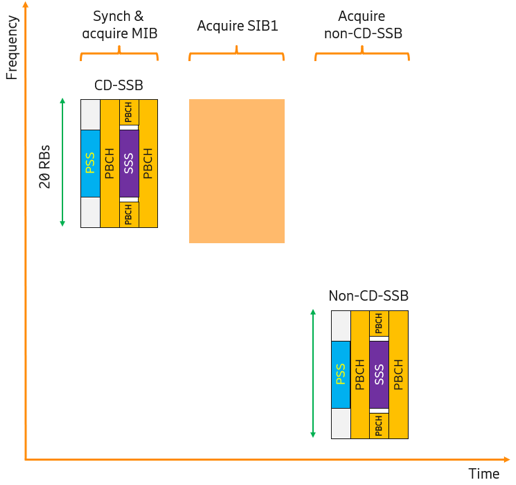}
\caption{An illustration of the UE procedure when utilizing two SSBs located in different frequency positions for the joint estimation of Doppler shift and frequency offset.}
\label{fig:2}
\end{figure}

\section{Simulation Results}
\label{sec:sim}

In this section, we provide link level simulation results to evaluate the Doppler shift estimation performance based on two SSBs located in different frequency positions. The simulation assumptions used are based on the ones outlined in [3], and are summarized in Table I.

\begin{table}
\centering
\caption{Link level simulation assumptions}
\begin{tabular}{|l||l|} \hline
Carrier frequency & 2 GHz  \\ \hline 
Subcarrier spacing & 30 kHz \\ \hline 
DL reference signal & Two SSBs  \\ \hline 
Channel Model	& NTN\_TDL\_D  \\ \hline 
Antenna configuration & 1 TX, 2 RX \\ \hline 
Frequency offset & [-10.5, 10.5] ppm \\ \hline 
Doppler shift & [-24.5, 24.5] ppm \\ \hline 
\end{tabular}
\label{tab:sys:para}
\end{table}

Figure 3 shows the cumulative distribution functions (CDF) of the Doppler shift estimation errors with different values of the frequency separation between the two SSBs ranging from 288 MHz to 2016 MHz. The signal-to-noise ratio (SNR) operating point is -3 dB. The results show that the larger the frequency separation between the two SSBs, the better the performance of the Doppler shift estimation, which aligns with expectation. Typically, a frequency error larger than 5\% of the subcarrier spacing would lead to noticeable inter-carrier interference. With 30 kHz subcarrier spacing and requiring the Doppler shift estimation error to be within 5\% of the subcarrier spacing, the Doppler shift estimation error should be within $[-1.5 \textrm{ kHz}, 1.5 \textrm{ kHz}]$. From Figure 3, we can see that with 864 MHz frequency separation between the two SSBs, more than 94\% of the Doppler shift estimation errors are within the range of $[-1.5 \textrm{ kHz}, 1.5 \textrm{ kHz}]$.

\begin{figure}
\centering
\includegraphics[width=7cm]{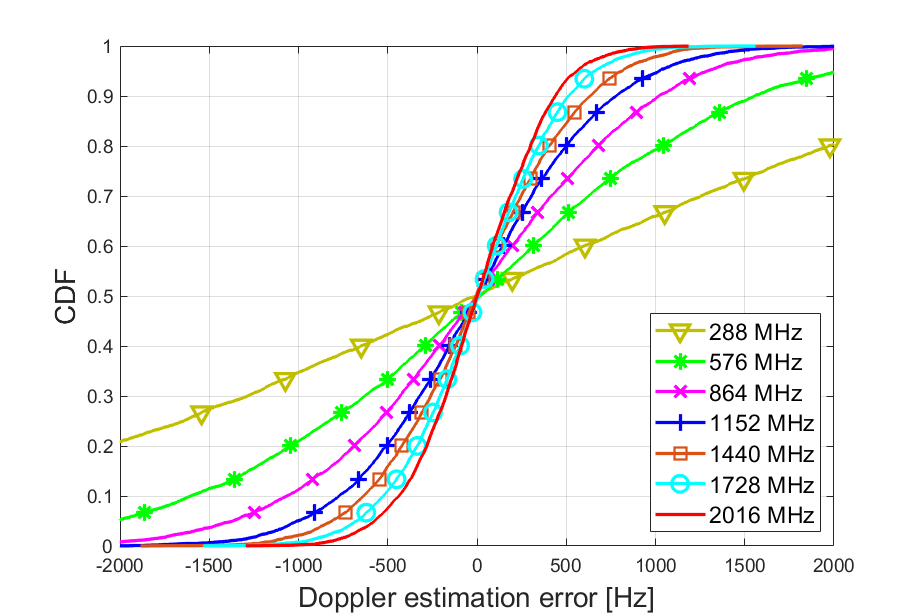}
\caption{CDF of Doppler estimation errors with different values of the frequency separation between the two SSBs under -3 dB SNR.}
\label{fig:3}
\end{figure}

Figure 4 shows the maximum and mean Doppler shift estimation errors versus the value of the frequency separation between the two SSBs under -3 dB SNR. Similar to the observation made for Figure 3, both the maximum and mean Doppler shift estimation errors decrease as the value of the frequency separation between the two SSBs increases. When the value of the frequency separation between the two SSBs becomes larger than about 400 MHz, the mean Doppler shift estimation error starts to be smaller than 1.5 kHz. 

\begin{figure}
\centering
\includegraphics[width=7cm]{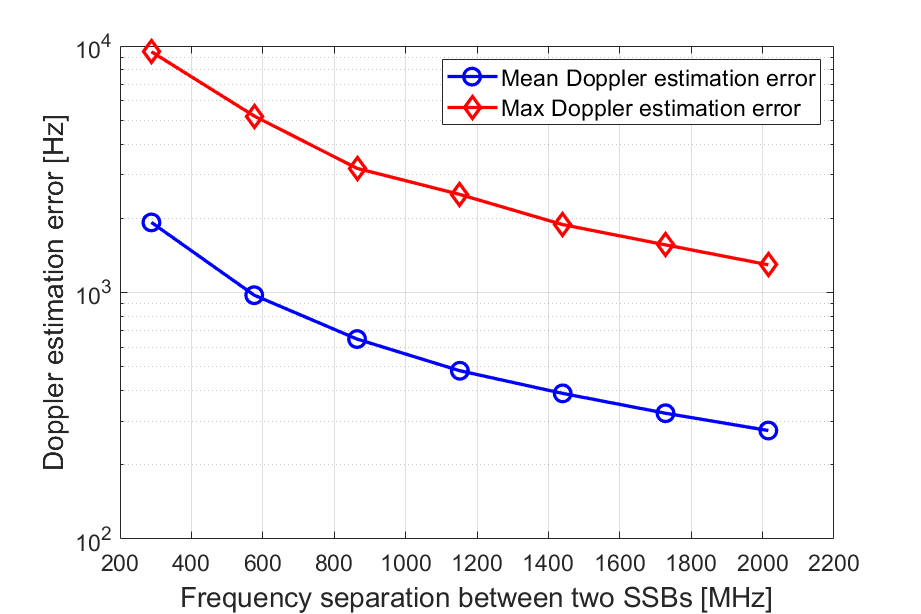}
\caption{Maximum and mean Doppler estimation errors versus the value of the frequency separation between the two SSBs under -3 dB SNR.}
\label{fig:4}
\end{figure}

Figure 5 shows the CDF of the Doppler shift estimation errors with different SNR values ranging from -3 dB to 13 dB. The value of the frequency separation between the two SSBs is fixed to be 288 MHz. As expected, the higher the SNR value, the better the performance of the Doppler shift estimation. From Figure 5, we can see that with 5 dB SNR, more than 90\% of the Doppler shift estimation errors are within the range of $[-1.5 \textrm{ kHz}, 1.5 \textrm{ kHz}]$.

\begin{figure}
\centering
\includegraphics[width=7cm]{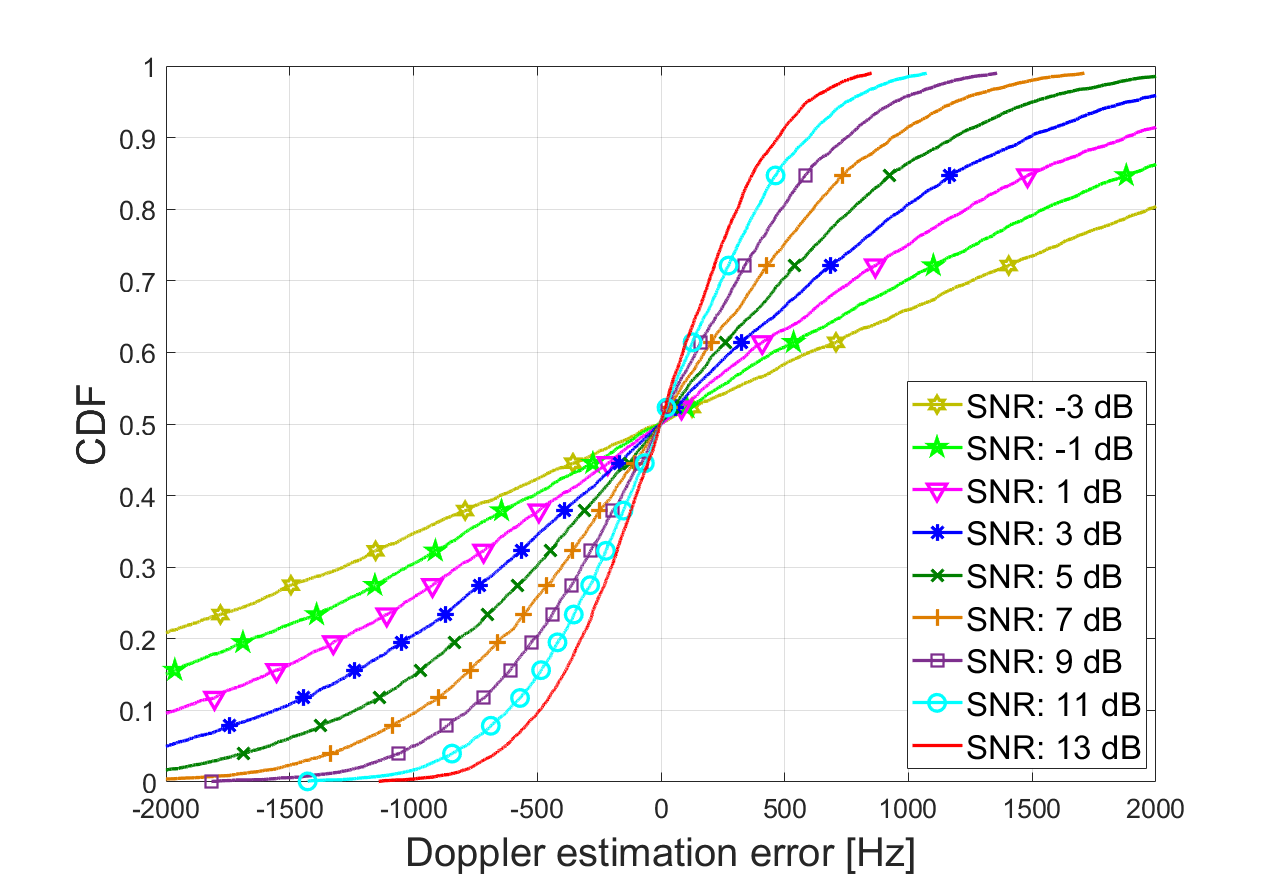}
\caption{CDF of Doppler shift estimation errors with different SNR operating points: 288 MHz frequency separation between the two SSBs.}
\label{fig:5}
\end{figure}

Figure 6 shows the maximum and mean Doppler shift estimation errors versus SNR. The value of the frequency separation between the two SSBs is fixed to be 288 MHz. Similar to the observation made for Figure 5, both the maximum and mean Doppler shift estimation errors decrease as the SNR value increases. When the SNR becomes larger than 0 dB, the mean Doppler shift estimation error starts to be smaller than 1.5 kHz. 

\begin{figure}
\centering
\includegraphics[width=7cm]{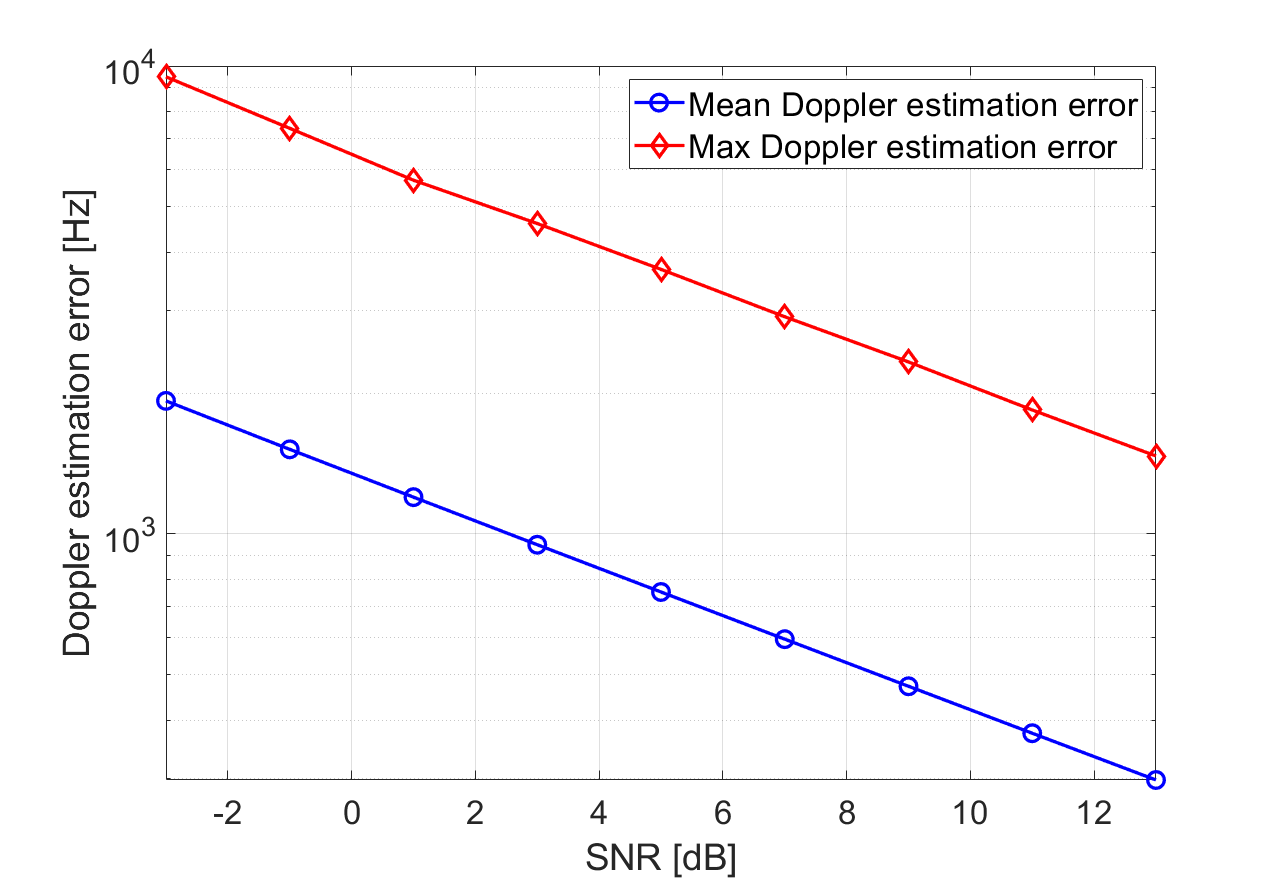}
\caption{Maximum and mean Doppler shift estimation errors versus SNR: 288 MHz frequency separation between the two SSBs.}
\label{fig:6}
\end{figure}

In summary, the simulation results validate the feasibility of the estimation of Doppler shift in the presence of frequency offset due to oscillator mismatch, by utilizing two SSBs located in different frequency positions. With the estimated Doppler shift, UE can determine the needed frequency adjustment value in its uplink. This can reduce the dependency of NTN operation on GNSS. To achieve better estimation accuracy of Doppler shift, one could either increase the frequency separation between the two SSBs or increase the SNR operating point. The possibility of increasing the frequency separation between the two SSBs mainly hinges on the spectrum allocated to the system. The possibility of increasing the SNR operating point depends on a number of factors such as onboard payload power budget, deployable antenna in space, and UE's antenna configuration. For example, more advanced user devices, such as very small aperture terminal (VSAT), can enjoy high SNR with their high gain receive antennas, which the handheld devices may not have.

\section{Conclusions}

In this paper, we propose to utilize reference signals in more than one frequency position in an OFDM carrier for an NTN device to jointly estimate the Doppler shift (mainly due to satellite movement) and frequency offset (due to crystal oscillator mismatch between the transmitter and the receiver). We explain the design rationale by describing the signal processing model and analysis, elaborate how to utilize the NR reference signals, and present simulation results. The proposed solution can reduce the dependency of NTN operation on GNSS.

Support of high mobility communications is a prominent feature of 5G networks. The design presented in this paper is not limited to NTN but is also applicable to other high mobility scenarios. One example is the high-speed train (HST) scenario, which is of special importance because of the fast expansion of HST systems worldwide and the great demands of high-speed connectivity from passengers and HST special services. The importance of HST scenario has triggered 3GPP to conduct study and work items to introduce enhancements to tackle the challenges associated with HST with speeds up to 500 km/h. Doppler effect is one of the main challenges in HST scenarios. The design presented in this paper can be used to estimate the Doppler shift in HST scenarios and thus can help to improve the HST connection performance.

%\bibliographystyle{IEEEtran}
%\bibliography{IEEEabrv,Reference}

\end{document}